# Using Static Charge on Pyroelectric Crystals to Produce Self–Focusing Electron and Ion Beams and Transport Through Tubes

James D. Brownridge[1] and Stephen M. Shafroth[2]

[1]Department of Physics, Applied Physics and Astronomy
State University of New York at Binghamton
Binghamton, New York 13902-6000, USA
e-mail: jdbjdb@binghamton.edu

[2]Department of Physics and Astronomy
Department of Physics, Applied Physics and Astronomy
University of North Carolina at Chapel Hill
Chapel Hill, North Carolina 27599-3255, USA
e-mail: shafroth@physics.unc.edu

***Abstract*- Static charge in and on the surface of pyroelectric crystals of $LiNbO_3$ and $LiTaO_3$ in a dilute gas has been shown to ionize gas molecules via electron tunneling. The released electrons and positive ions are focused and accelerated according to the sign of the static uncompensated charge. The uncompensated charge is produced when the temperature of the crystal is changed from any initial temperature between about 500K and about 15K. It may be either polarization charge that is inside the crystal surface or compensation charge that is on the surface. The direction of temperature change and the polarity of the base of the crystal determine whether electrons or positive ions are accelerated toward or away from the crystal. The ionization, focusing and acceleration may continue for more than 15 days following a single change in temperature from ~20 $^oC$ to ~160 $^oC$ to ~20 $^oC$.**

## I. INTRODUCTION

Twenty-three centuries ago the Greek philosopher, Theophrastus, described what is now call the "pryoelectric effect". He described how the mineral tourmaline had the power to attract objects such as straw, wood, copper and iron [1]. The phenomenon that allowed tourmaline to attract these materials is the spontaneous change in polarization in tourmaline in response to changes in temperature in either direction, of the mineral. We now know that tourmaline belongs to a large class of materials (often crystals) call pyroelectrics, these are materials that spontaneously polarize when their temperatures fall below their Curie temperatures. They remain polarized as long as they are in an environment where the temperature is below their Curie point [2, 3]. However, manifestation of polarization by the presence of electric charge at the surface is usually masked by the collection of compensating charge from the environment, if the material remains at a stable temperature for a short time. That is why Theophrastus had to change the temperature of tourmaline in some manner to cause it attract other objects.

It was not until 1756 that Franz Ulrich Theodor Aepinus undertook the first scientific study of the phenomenon. He found among other things that tourmaline would become electrically charged when heated and that one end would always be negative on rising temperature and the other would be positive. He also observed that this was due to the internal structure [1]. This was a key observation. However, how the internal structure caused charge to appear at the surface of pyroelectric crystals remained a mystery until the crystallographic structure of lithium tantalate ($LiTaO_3$) was determined in 1972 by Abrahams et al [2]. In 1985 Weis and Gaylord [3] summarized the physical properties and crystal structure of lithium niobate ($LiNbO_3$) to show that it is the movement of the Li and Nb atoms with respect to the O atoms in lithium niobate that causes the polarization to change with temperature. In lithium tantalate it is the movement of the Li and Ta atoms that cause the polarization to change with temperature. It is assumed that a similar process occurs in the mineral tourmaline.

Now, more than twenty-three centuries after Theophrastus we are still observing new phenomena associated with changing the temperature of pyroelectric crystals [4–9].

## II. STATIC CHARGE

By static charge we mean charge that will linger in or on the surface of a crystal for up to 15 days. Following in the traditions of the ancient Greeks we simply change the temperature of a pyroelectric crystal of lithium niobate ($LiNbO_3$) or lithium tantalate ($LiTaO_3$) instead of tourmaline to produce static charge in and on the surface of these crystals. To ensure that the charge is stable and long lasting we carefully control the environment, the rate of change in temperature and the range over which the temperature is changed. In Fig. 1 we show a $LiNbO_3$ crystal attached to a heater in a vacuum chamber. The heater is a 62–ohm resistor and is used to change the temperature of the crystal. The vacuum chamber is used to control the environment around the crystal. By controlling the environment (gas pressure) we control the availability of compensating charge.

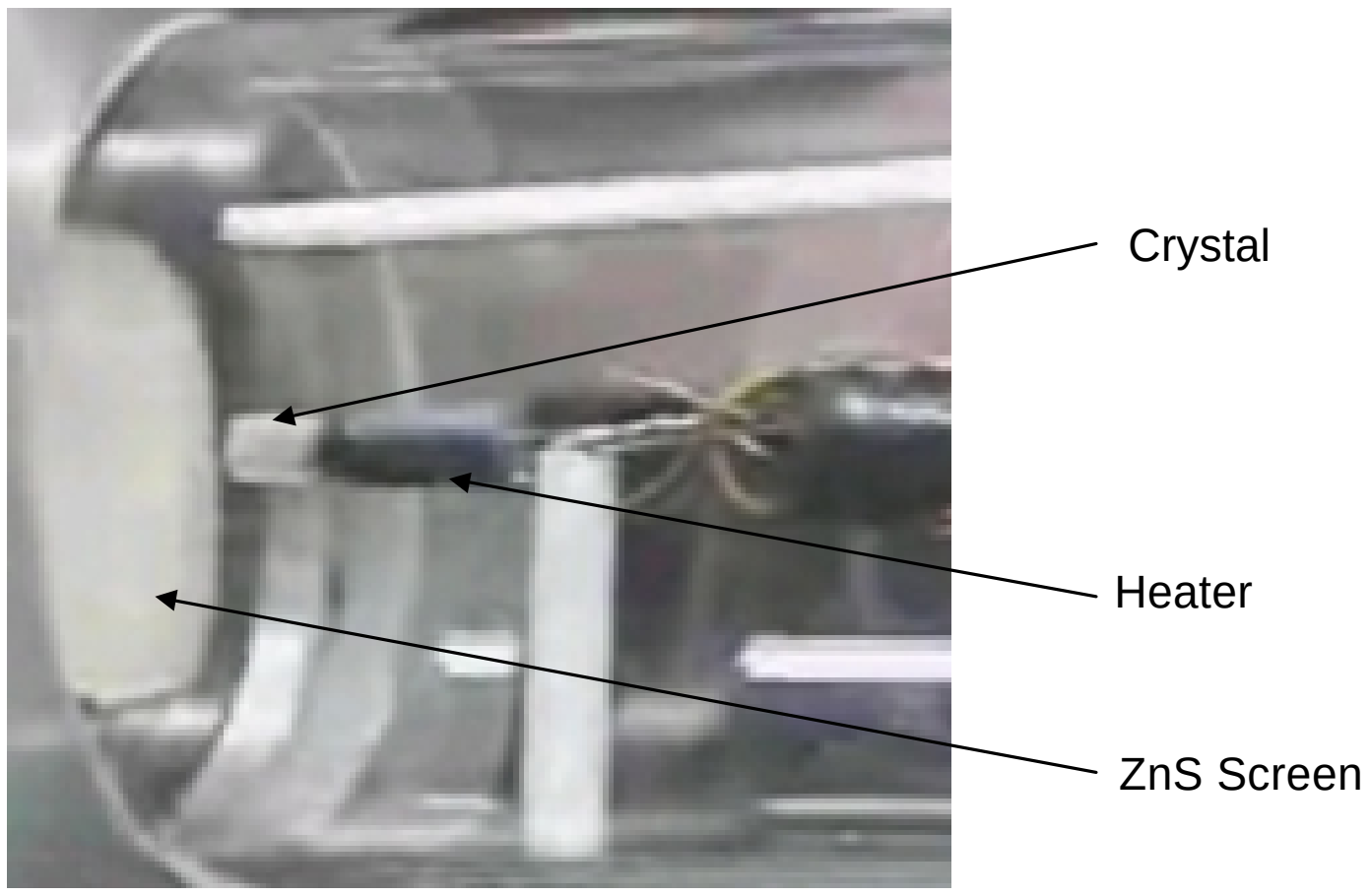

Fig. 1. Photograph of the end of the vacuum chamber that houses the crystal, crystal heater and the ZnS screen. The -z end of the crystal is at a distance of 21 mm from the ZnS screen. The crystal is $LiNbO_3$ 10 mm long and ~5 mm in diameter.

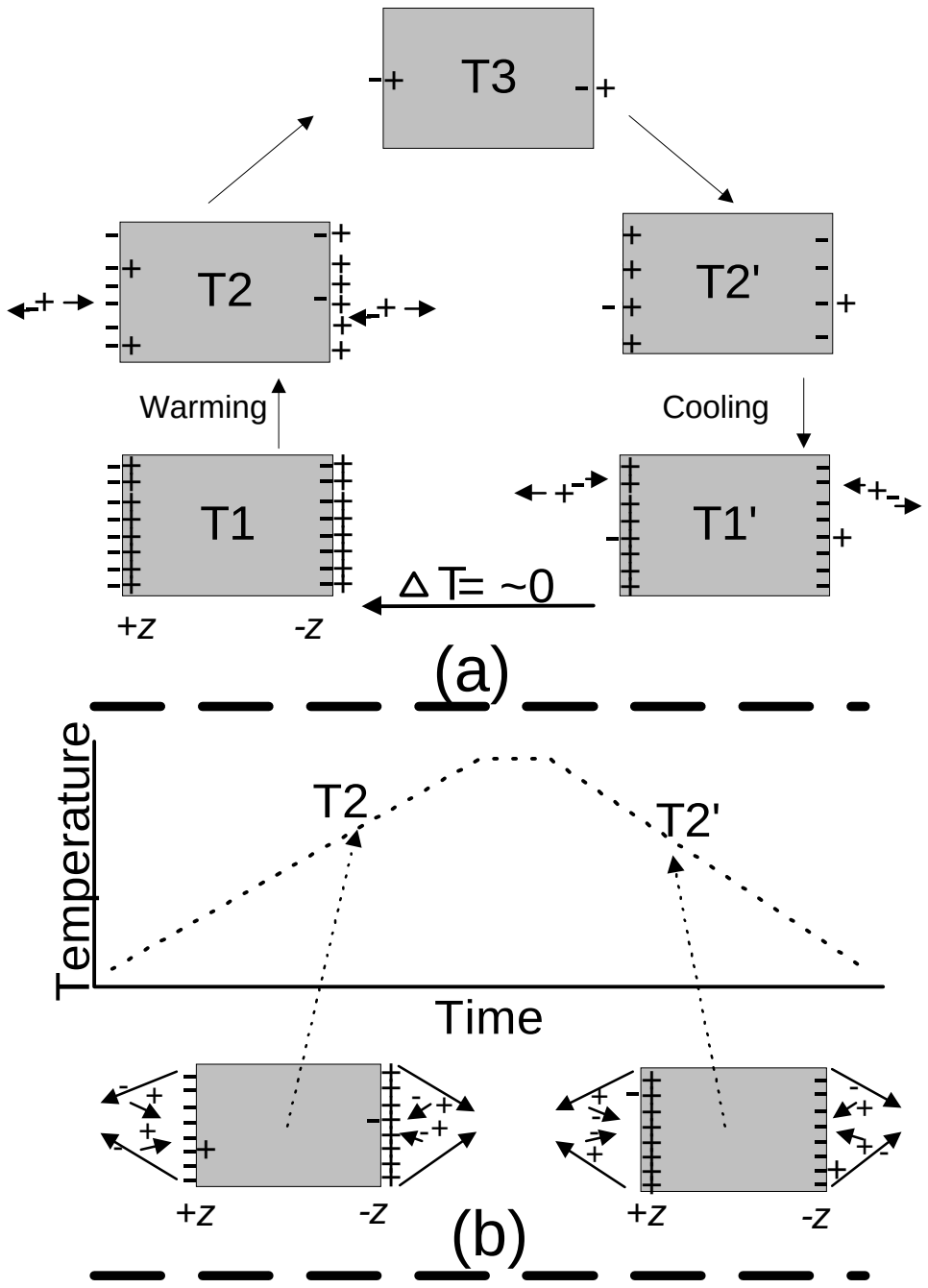

Fig. 2. Schematic diagram to provide a descriptive picture of the response to temperature changes of a pyroelectric crystal in a dilute gas environment. In an operating system one end of the crystal is always in contact with the heater.

In an environment of room air at 1 atm charge from the air surrounding the crystal will compensate or neutralize the polarization and surface charge almost as fast as it is produced and at the end of a temperature change the net charge on the surface of the crystal will be zero. In Fig. 2(a) we show schematically an idealized version of how we arrive at a state of "static charge" as defined above in/on the surface of a crystal. We began at a temperature T1 in Fig. 2(a) where all polarization charge inside the crystal is compensated for by charge of opposite sign attracted to the surface of the crystal. As the temperature of the crystal is raised from T1 through T2 towards T3 the internal polarization charge is reduced at a rate proportional to the rate of change in temperature and the pyroelectric coefficient of the crystal. This leaves uncompensated external charge on the outer surface that builds up at the same rate that polarization charge is reducing. This external charge may be physically removed in one discharge if it builds up too fast or in most cases it is neutralized via ionization of residual gas molecules in the vacuum chamber. The rate of ionization is pressure dependent. When the temperature is returned to T1' which is approximately equal to the original temperature T1, there will be a large uncompensated polarization charge at the internal surface of the crystal. This is the "static charge" that will best focus and accelerate electrons and positive ions. The "static charge" that is produced on rising temperature will also focus and accelerate ions and electrons but it is not as stable because it can be easily dislodged

## III. SELF–FOCUSING

Self-focusing of positive ions and electrons into beams [4] are two of several phenomena observed when the temperature of a crystal of $LiNbO_3$ or $LiTaO_3$ is cycled through ~100 $^oC$ in a dilute gas environment. The exact magnitude of the temperature change is not important. The crystal must be cut perpendicular to its threefold rotationally symmetric z-axis. In Fig. 2b we show a schematic cartoon (a snapshot in time) of the behavior of positive ions and electrons at two temperatures, T2 and T2' as the crystal is cycled from ~20 $^oC$ to ~120 $^oC$ to ~20 $^oC$. On rising temperature, at the *-z* base positive ions are accelerated away from the crystal in a focused beam and simultaneously electrons are accelerated to it in a focused beam striking near its center (see Fig 3b). When the crystal is cooling the process is reversed as shown schematically in Fig. 2(b) at T2'. The surface of the crystal was covered with a thin layer of ZnS. (c) Shows electron beam image when electron are accelerate away from a $LiTaO_3$ crystal and strike a ZnS screen.

If both ends were exposed these four processes would occur simultaneously. When electrons strike the crystal they cause the production of K x rays of Nb when the crystal is $LiNbO_3$ and L x rays of Ta when the crystal is $LiTaO_3$ [5]. In Fig. 3(a) we present a photograph (snapshot) showing the crystal. We imaged the spot where electrons struck the crystal by covering the surface of the crystal with a thin layer of ZnS in a clear lacquer. Zinc and S K x rays are produced when the spot is visible. The system is symmetric and we see the reverse of this process on cooling. In Fig. 3(c) we show the evidence of electrons accelerated away from the crystal in a focused beam. By placing a ZnS screen (see Fig. 1) at the focal length of the crystal we were able to image the beam spot [4, 6]. We have not observed the four events simultaneously because the experimental arrangements that we use require that we heat the crystals from one end and observe the other. However, we are able to observe simultaneously two events at each end of the crystal during separate runs.

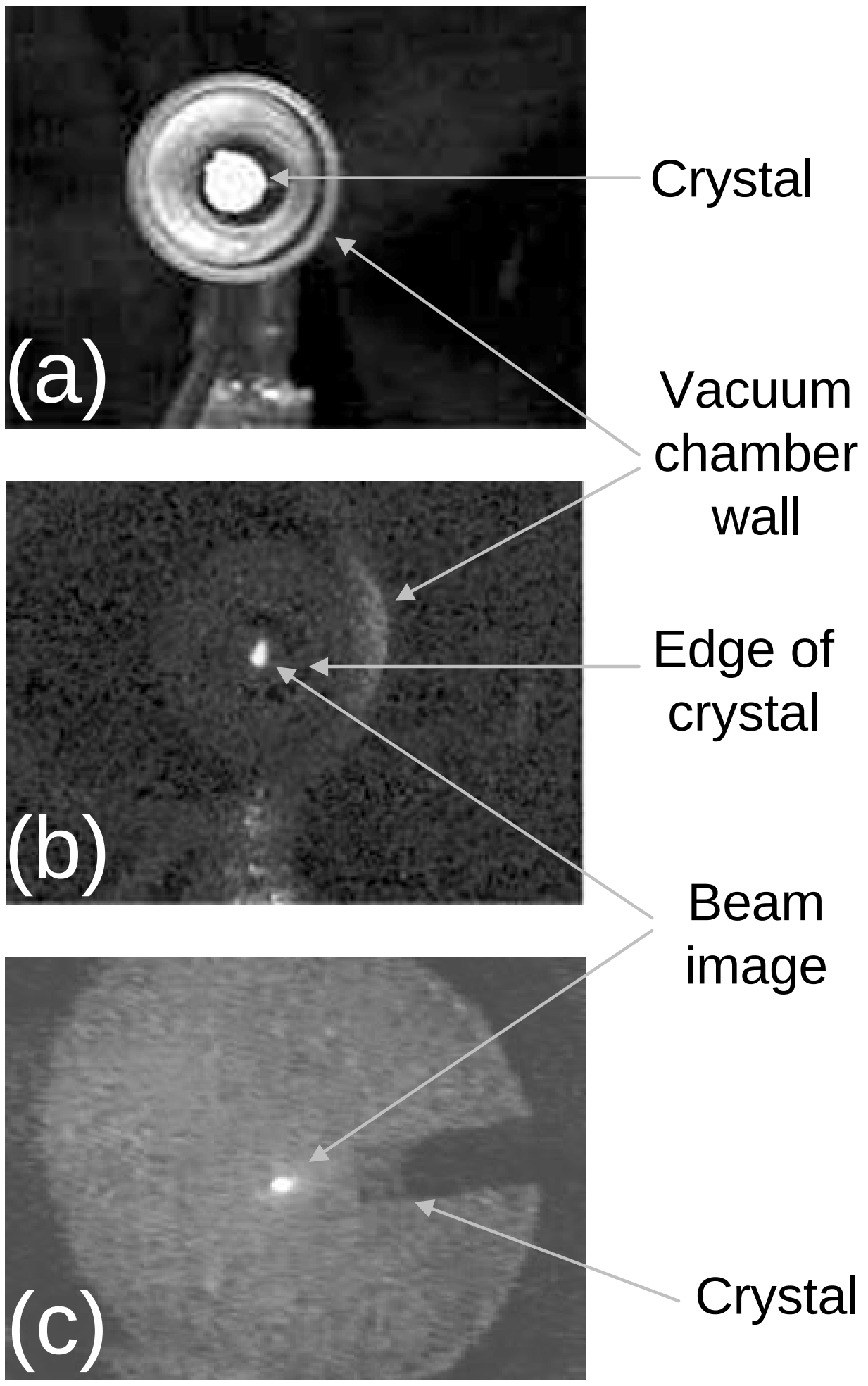

Fig.3. (a) A 6 mm diameter by 10 mm $LiNbO_3$ crystal is shown, end view, in a glass vacuum chamber. (b) Shows the fluorescent spot where electrons strike the crystal when they are accelerated to it. The surface of the crystal was covered with a thin layer of ZnS. (c) Shows electron beam image when electron are accelerate away from a $LiTaO_3$ crystal and strike a ZnS screen.

We have not been able to image the positive ion beam. However, we have measured some of its characteristics with the surface barrier detector and slit that was used to scan the beam. Details of electron beam characterization can be found in references [4, 6–9]. In brief, we scanned across the positive ion beam at various distances from the crystal with the electron detector and determined that ions accelerated away from the +*z* base on cooling were focused to a spot some distance away from the crystal. It is assumed that positive ions accelerated to the crystal are also focused to a spot near the center of the crystal.

We believe that positive ions and electrons are focused because the electric field lines near the crystal surface are slightly inclined toward the *z*-axis due to the higher electric field at the edge of the crystal relative to that at the center.

## IV. SOURCE OF POSITIVE IONS AND ELECTRONS

The source of positive ions and electrons that makes up these beams is the dilute gas in the vacuum chamber. Because of the static charge in and/or on the surface of the crystal, there is a very strong electric field at the surface of the crystal ($|E_0| = 1.35\text{x}10^7$ V/cm) [10]. Thus, when gas molecules randomly approach the exposed surface of the crystal they are ionized via electron tunneling, thus producing an ion pair. One specie is accelerated away and the other towards the crystal. They are both focused by the same electric field that caused their production. The species that is accelerated to the crystal compensates or neutralizes static charge on or in the surface of the crystal. This causes a reduction in the field strength of the electric field. Over time the strength of the electric field will be reduced to below the threshold for ion pair production. The pressure in the vacuum chamber determines the time it takes for the electric field to fall below the threshold for ion pair production. At pressures < $10^{-6}$ Torr we have observed ion pair production for 15 days following one temperature cycle [9].

## V. BEAM TRANSPORT

Positive ion and electron beams have been transported through small diameter metal and plastic tubes ranging in length from 21.5mm to118mm with inside diameters from 0.25mm to 0.75mm. The tubes were both straight and inclined with the beams, or slightly curved with the entrance in line with the beam, or straight and at a slight angle to the beam or the beams focal spot was offset from the entry to a straight tube. In each case the tubes opening was at the focal length [4]. In Fig. 4 we show schematic drawings of two of the tube arrangements described above. In Fig. 5 we show a positive ion and three electron spectra collected using the arrangement shown in Fig. 4a. Figure 5a is the positive ion spectrum collected during the 156 sec. warm-up. Figures 5b-d shows electron spectra collected as snapshots as the crystal cools in three different pressure environments. Notice that the energy of the first peak increases with increasing pressure a phenomenon we call gas amplification. Gas amplification and the multi-peaks are discussed in detail in references 7 and 8.The electron spectra which show peaks at energy E, 2E, 3E… arise because when the exposed crystal surface is negative positive ions bombard it and eject several electrons in one collision due to Auger electron emission and secondary electrons. These electrons are focused and arrive at the detector within its resolving time. Thus there are no electrons of energy 2E, 3E …Here we point out that transport of the electron beam through metal or plastic tubes does not impede these processes when the tube is straight. On the other hand as shown in Fig. 6 when the electron beam passes through the curved tube as shown in Fig. 4(b) the multi-peak information is lost. However, gas amplification and count rate increase with pressure is still evident. The electron count rate increase is due to increased available gas molecules for ionization during each 10 sec. counting interval.

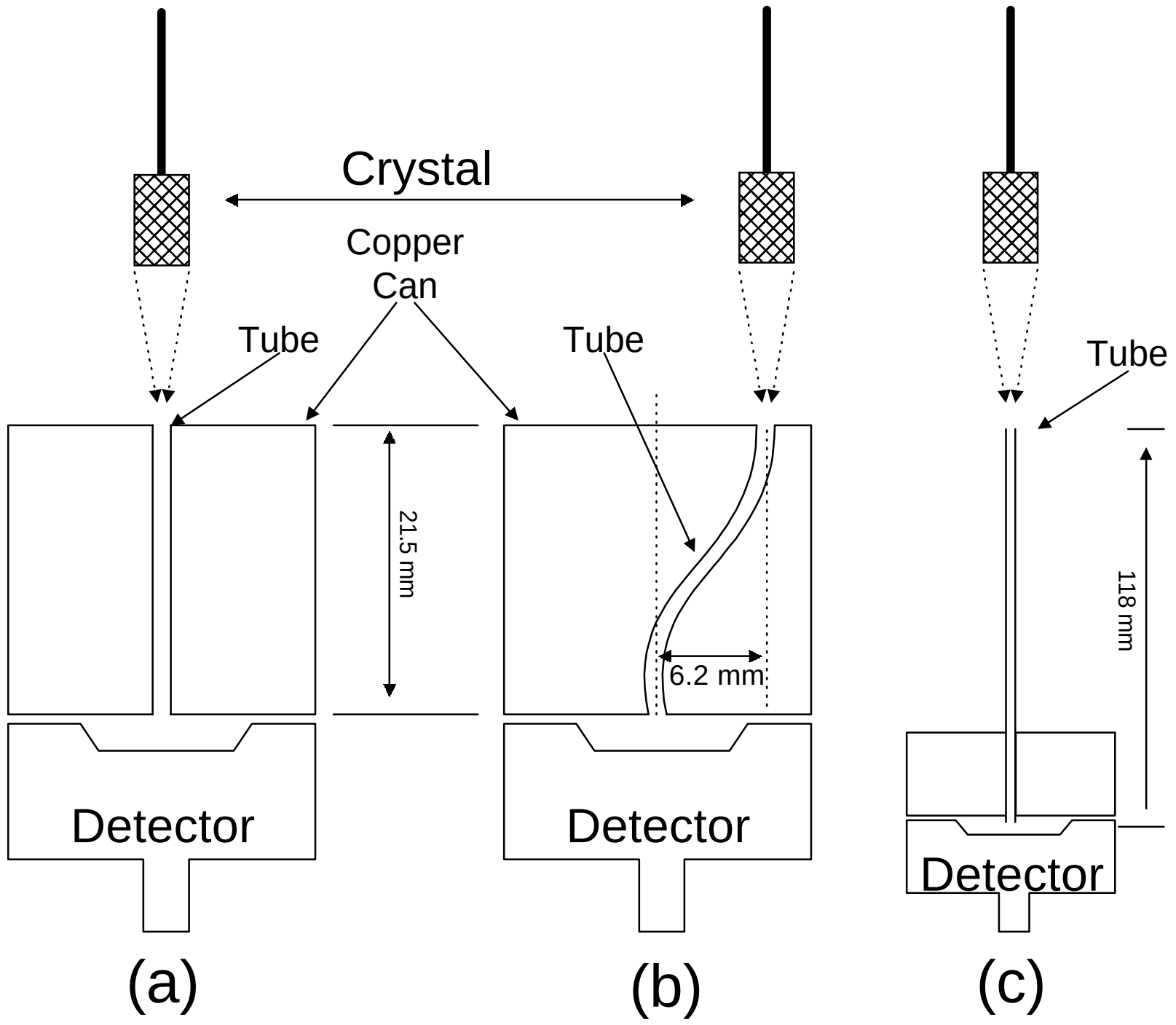

Fig. 4. The detector is an energy sensitive surface barrier electron detector. The crystal is $LiNbO_3$ 10mm long and ~5mm in diameter. The curved tube is plastic and the straight tube is metal. The tubes are interchangeable between the copper metal cans. The copper can-detector assembly can be moved with respect to the spot where the beams are focused.

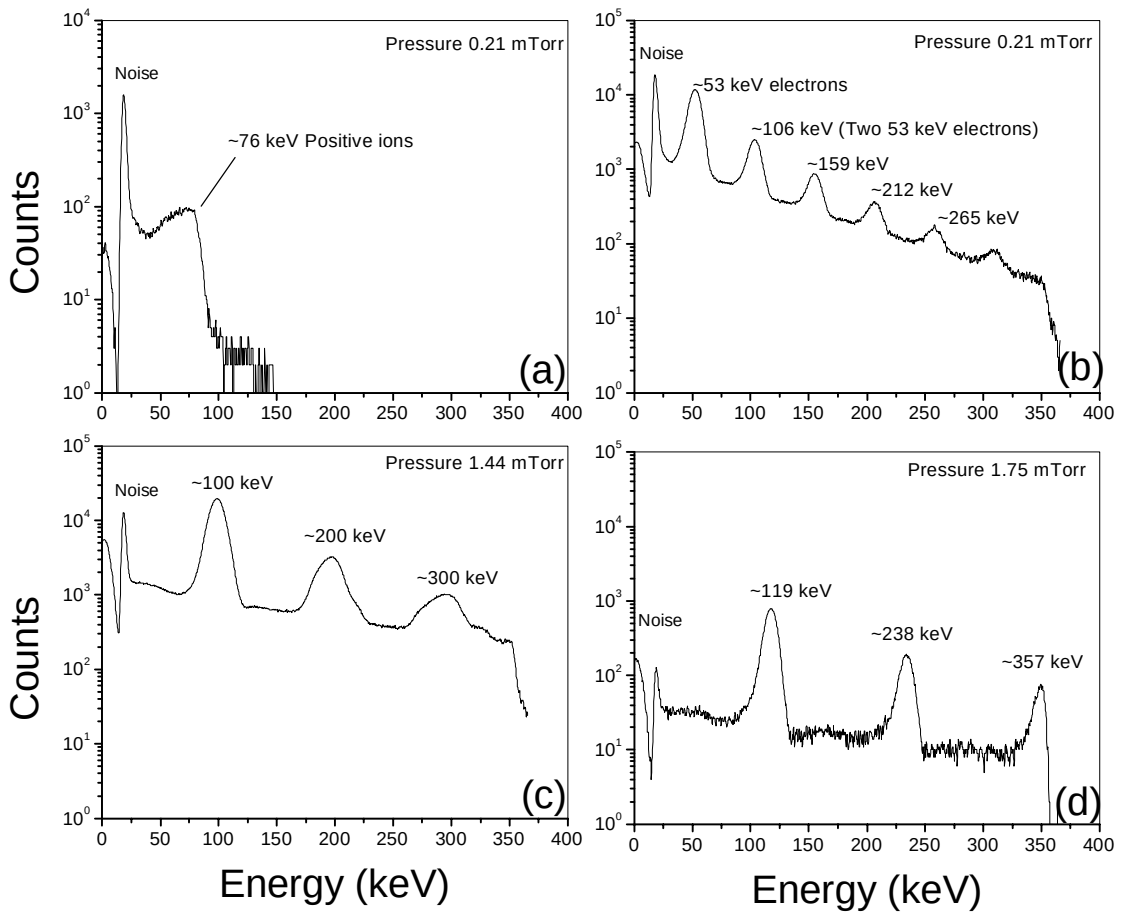

Fig. 5. (a) Positive ion spectra collected during crystal warm-up using the arrangement shown in Fig. 4a. (b-d) Electron spectrum collected using the same arrangement but on cooling and at different gas pressures. The multiple electron energy peaks are the results of two, three, four, etc., electron arriving at the detector within the resolving times of the electronics.

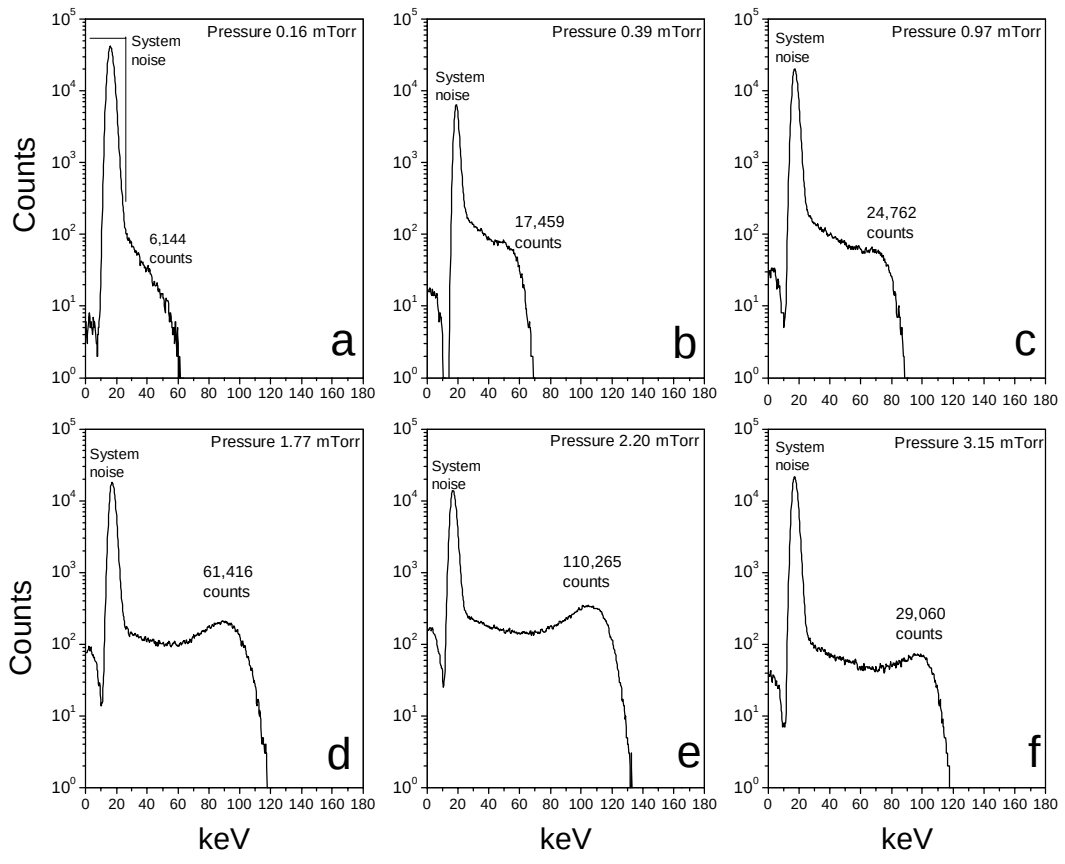

Fig. 6. Electron spectrum observed when the arrangement is as shown in Fig. 4(b). Each spectrum was collected at a different gas pressure over about the same temperature range (~70 $^{o}$C to ~28 $^{o}$C) and for the same length of time (1,000 sec).

We have used several techniques to try and increase our understanding of the nature of the focusing process and the characteristic of the electron beam and its dependence on the environment around the focal spot. The arrangement shown in Fig. 4(c) was used to determine if focusing would be disturbed or distorted by placing a small object at or near the focal spot. A metal tube with an inside diameter of 0.75 mm was used. The tube could be grounded and ungrounded during experiments. No effect was observed.

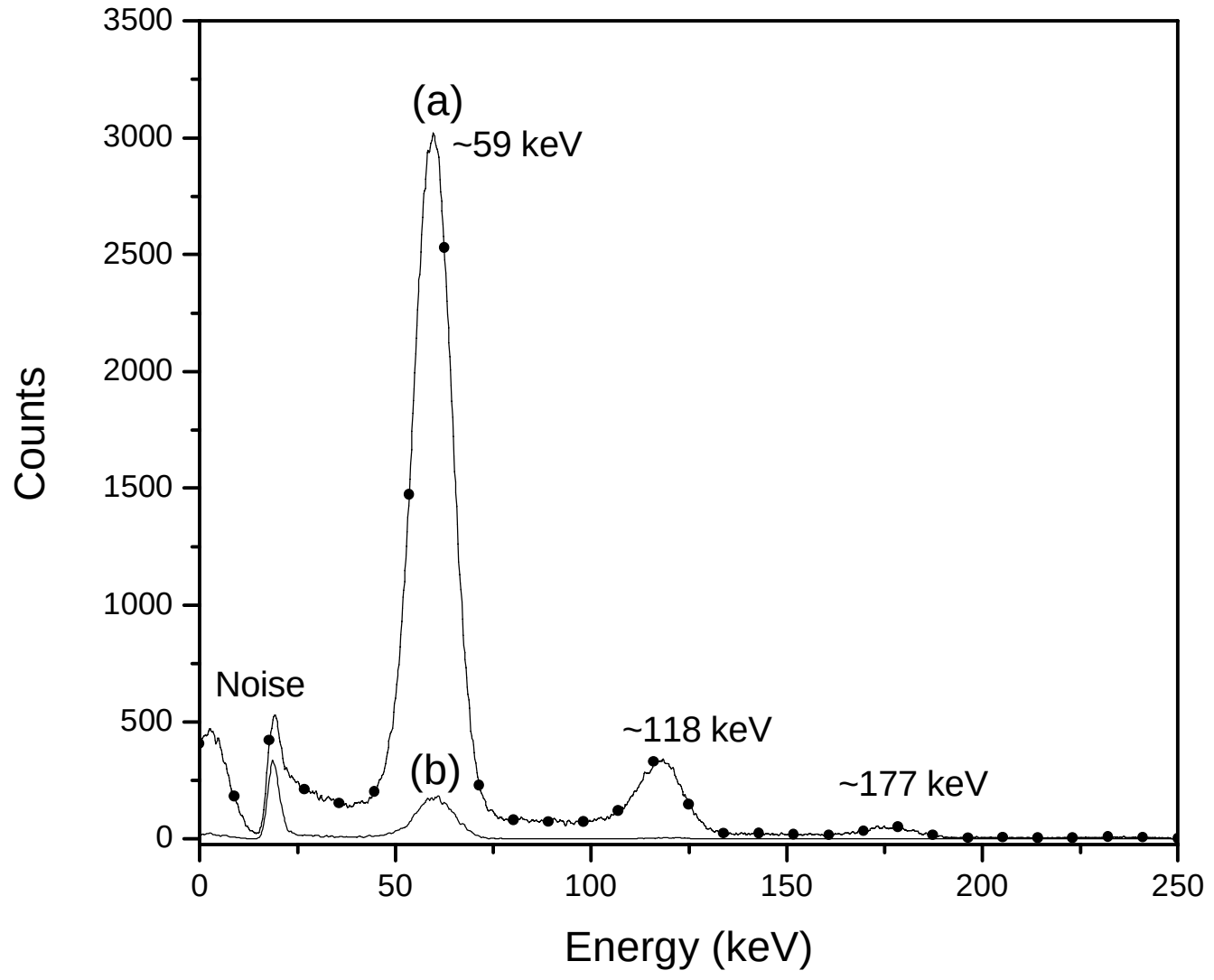

Fig. 7. Electron spectrum observed when the arrangement is as shown in Fig. 4(c). Spectrum (a) was collected when the opening of the 118 mm long metal tube as at the focal spot and the beam went down the tube to the detector. Spectrum (b) was collected when the tube was moved to one side by ~2mm. The counting time for each spectrum was 10 seconds.

Evidence that the tube does not appear to affect the focusing is presented in Fig. 7. Here we show electron spectrum collected when the tube's opening is at the focal spot (a) and when it is moved ~2mm away to one side (b). The tube was moved back and forth many times during cooling with reproducible results.

High-count rates with multiple peaks were observed when the opening was at the focal spot and low counting rate when it was moved away. Moving it several more mm resulted in no counts.

ACKNOWLEDGMENT

We are most grateful to our colleagues, Sol Raboy, Eugene Merzbacher, and Tom Clegg for many insightful discussions and continued encouragement.